A Sommerfeld Explanation
Jeremy Bernstein
Stevens Institute of Technology

Abstract: We discuss how the Wien displacement law reflects the quantum theory.

Periodically I prepare an imaginary lecture the purpose of which is to remind philosophers of physics, and indeed some physicists, that the quantum theory had its origins in experiments. Unlike general relativity which seems to have sprung from Einstein's head, the quantum theory was a response to experiment. Even de Broglie's conjecture that particles had also a wave-like nature was influenced by how this notion could be used to explain the quantization of the radii of the Bohr orbits. The experiment driven theoretical developments of the quantum theory began with Planck and have continued ever since. In my most recent preparation I decided to re-read Thomas Kuhn's book Black- body Theory and the Quantum Discontinuity.1894-1912 which was first published in 1978.[1] But his black-body book is a splendid excursion into the history of physics.

Kuhn sets the stage for Planck. In the winter of 859-60 the German physicist Gustav Kirchoff laid down the dimensions of the problem. He considered a cavity whose walls were heated to a temperature T. The walls emitted and absorbed radiation and in equilibrium this radiation acquired the same temperature. He then defined $a_\lambda$ and $e_\lambda$ as the fractions of the radiation absorbed or emitted in the wave length range $\lambda + d\lambda$. He argued that for all materials the ratio $e_\lambda / a_\lambda = K_\lambda(T)$ must be a universal function of these two variables. A black-body is one in which $a_\lambda = 1$. This means that measuring the spectrum of the emitted radiation can be used to determine K.

In 1884 Boltzmann made a contribution to the subject of black-body radiation which was important because the German physicist Wilhelm Wien made use of his method when he demonstrated what is really the main subject of this little note-the Wien displacement law. What Boltzmann did was to imagine a cylinder of volume V closed by a piston. The radiation in V is assumed to exert a pressure on the piston. If V is increased in such a way as to keep the temperature constant then heat must be added. Let the energy density be u and the volume V so that the total energy $U=uV$. Boltzmann knew that the radiation pressure p was $1/3u$. Hence the first law of thermodynamics reads

$$\delta Q = V \partial u/\partial T \, \delta T + (V \partial u/\partial V + 4/3 u)\, \delta V. \qquad (1)$$

Boltzmann had shown that u and K must be proportional each other in equilibrium and therefore u which is an integral over all wave lengths can like K be a function of temperature alone. We now proceed as follows. From the work of Clausius who defined the notion of entropy it was known that
$\delta S/T = \delta Q/T$ is an exact differential so that
$\partial^2 S/\partial T \partial V = \partial^2 S/\partial V \partial T$. Using the fact that u is independent of V we can derive the ordinary differential equation
$du/dT = 4u/T$ which has the solution $u = \sigma T^4$. The job was then to find $u(\lambda, T)$. Enter Wilhelm Wien.

Wien, who won the Nobel Prize in physics 1n 1911 was born in East Prussia in 1864. His father was a land owner and Wein was heading in the direction of becoming gentleman farmer when during his school years he began studying physics and mathematics. He took his PhD from Hermann von Helmholtz. His thesis was primarily experimental and indeed his career was both as a theorist and experimenter. It was in Helmholtz's laboratory in the early 1890's that he did his important work on black-radiation. What concerns me here is his "law of displacement" which says that $K_\lambda(T)$ must be proportional to $\lambda^{-5} \varphi(\lambda T)$. The power of $\lambda$ is clear given the form of $\varphi$. We must upon integration over $\lambda$ recover Boltzmann's $T^4$ But how did two variables contract into one? This is exactly the point where I got stuck the first time I read Kuhn's book and I am reading it a second time and am stuck again .Kuhn is no help whatsoever. This genuinely annoys me but now I am determined to something about it.

In the day I taught thermodynamics and statistical mechanics a couple of times. We used Fundamentals of Statistical and Thermal Physics by Fred Reif[2]. It is a good solid text and I certainly used it when I discussed black-body radiation. I still have a copy so I looked to see what he had to say about the displacement theorem. It is of course there but it appears as a consequence of the Planck distribution. This is of course the exact reverse of the historical order. Planck made use of Wien's theorem in the discovery of his distribution. It simplified his task enormously since he was then looking for a function of only a single variable. For my purposes Reif's treatment was useless. I had a couple of other texts that did the same thing.

In desperation I hit the web. If you type in "Wien's displacement law" you will be overwhelmed. It is like trying to get a drink of water from a fire hydrant. However occasionally you hit a pool of ambrosia. In this case it was a web site called bado-shanai.net/map of physics/mopWienslaw. I wish I could tell you more about this site such as who is responsible for it. What I can tell you is that the essays I have so far downloaded from "map of physics" have been uniformly excellent. You will not find a better essay anywhere, including Kuhn's book, on the steps Planck took in the discovery of his distribution law. The derivation of Wien's law is also very clear and the claim is made that it is essentially what Wien did. I am not going to reproduce it here because I am interested I telling you about another derivation which I found remarkable. The Wien derivation uses the notion of a spherical cavity of volume V filled with black body radiation. The cavity is allowed to expand adiabatically so that entropy is conserved. The derivation uses an expression for the entropy $S=4/3\sigma VT^3$ which is derived at some length in another "map." Thus if R is the radius of the sphere then during the expansion RT=constant. (Cosmological readers will find this a familiar equation.}There is then a geometrical argument involving the Doppler shift which leads to the equation
$d\lambda/\lambda = dR/R$ which implies that $\lambda_1/\lambda_2 = R_1/R_2$. Using the connection between R and T we have the Wein law. This is the outline and as they say the devil is in the details and the details are not so trivial.

I looked at several thermodynamics texts with similar results when I made a serendipitous discovery . It wasThermodynamics and Statistical Mechanics by Arnold Sommerfeld. The version I have was published in English by Academic Press in 1956.[3] Sommerfeld was one of the greatest teachers of physics who ever lived. His students included Heisenberg,Pauli, Bethe and Peierls. His lectures were famous. This set was published after his death in 1951. He died at age 82 after being hit by a truck while walking with his grandchildren. He had become deaf and did not hear the truck. He was still working on this volume of his lectures which was readied for publication by some of his colleagues in Munich. I thought that he would have something interesting about the displacement law and I was not disappointed.

He begins his discussion by saying what he will not do. He is not going to present Wien's argument. He gives reference to a 1945 paper by von Laue which he says contains the simplest proof along these lines. I have not seen this paper but it would not surprise me if this was true. What he discusses with whether the result can be made plausible by dimensional analysis alone. He gives reference to a paper by Glaser and Sitzugnsber where he says something similar was done.[4] I have no idea what these people did and have no access to the Viennese journal where their paper was published. In any event here is how Sommerfeld argues.

He notes that there are four fundamental units in classical thermodynamics exclusive of the electric charge which he says is irrelevant to these considerations. He calls u the energy density of the radiation per unit frequency. Hence giving the energy the symbol e the dimensions of u are $et/l^3$ where t is the time and l the length. A frequency v has dimensions $1/t$ and a temperature has its own dimension T. The speed of light has the dimensions of $l/t$ and Boltzmann's constant has the dimensions of $e/T$. These are the players in the game. We wish to find a dimensionless constant C made up out of the players. We take the exponent of u to be one and find that apart from a numerical coefficient which the dimensional argument cannot reveal that $C=uc^3/v^2kT$. This tells us that
$u \sim v^2kT/c^3$. Hence we are led to the Rayleigh-Jeans form of the spectrum with its attendant ultra-violet catastrophe. It should be noted that by the time Wien was doing his work such data as there was, was consistent with the spectrum he proposed which was proportional to $1/\lambda^5 \exp(-a/\lambda T)$, with 'a' a parameter that could be chosen to fit the data. This form was consistent with the displacement law and also the $T^4$ character of the integrated spectrum. It was Einstein who emphasized that such an expression could never arise from classical physics which is what the dimensional argument shows clearly. Incidentally a reader who is bothered by our choice of the dimensional exponent of u in the argument is invited to take another exponent and see if anything essential changes.

To make progress we must assume that there is another fundamental unit which has not been taken into account and which does not show up in classical physics. This new unit will allow us to construct a new constant C'. Sommerfeld notes that C' can be chosen so that it does not depend on u. If it did we could multiply it by some suitable power of C which would divide out u and a define a third constant which was independent of u. Having done this we could take powers of this constant so that the frequency occurs only to the first power. We might as well assume that C' has this property from the beginning. Hence

$$C' = dvT^n. \qquad (2)$$

The dimensions of 'd' will depend on the power n. Thus we can write

$$u(v,T) = v^2/c^3 kT \, F(dvT^n) \qquad (3)$$

The power n can be determined by the requirement that integrating u over all frequencies will produce an answer that goes as $T^4$. Let us call $x = dvT^n$. Then

$$u = kT^{1-3n}/d^3 c^3 \int_0^\infty f(x) x^2 dx. \qquad (4)$$

Thus to get the correct $T^4$ we must chose n=-1. Hence

$$u = v^2 kT/c^3 \, f(dv/T). \qquad (5)$$

This is the Wien displacement law. If we write d=h/k we have Planck's version. I do not know about you, but I find it remarkable that this humble displacement law that we all learned in freshman physics opens up a new universe. Planck thought so. Soon after he made this discovery in 1900 he went for a walk with his then seven year old son Erwin. This is the same Erwin Planck who was brutally executed by the Nazi's in January of 1945 for his part in the failed July 20,1944 plot to execute Hitler. Over the years Erwin said that during that walk he father told him that he had made the greatest discovery since Newton. Some times Erwin would say it was since Copernicus. Planck was sure that he had done something of great importance-and he was right.

---

[1] A new edition was published 1987 by the Chicago University Press. It is this edition that I will refer to. Thomas S. Kuhn, *Black- body Theory and the Quantum Discontinuity.1894-1912*, (University of Chicago Press. Chicago, 1987. Kurh supplies some seventy pages of notes for the reader who wants to explore the subject mpre deeply/

[2] Fred Reif, *Fundamentals of Statistical and Thermal Physics* ( McGraw Hill,New York) 1965.

[3] Sommerfeld.A, *Thermodynamics and Statistical Mechsnics*, (Academic Press, New Yor),1947)

[4] W.Glaser andJ.Sitzungsber, d Akad,Wien,Vol.156,**87**